\documentclass{article}

\usepackage{a4wide}
\usepackage{microtype}
\usepackage{graphicx}
\usepackage{amsmath,amssymb}
\usepackage{mathtools}
\usepackage[hidelinks]{hyperref}
\usepackage{booktabs}
\usepackage{tikz}
	\usetikzlibrary{external}
\usepackage{xcolor, colortbl}
	\newcolumntype{a}{>{\color{gray}}r}
\usepackage{cleveref}

\newcommand{\vi}{\textnormal{VI}}
\newcommand{\ri}{\textnormal{RI}}
\newcommand{\tj}{\textnormal{TJ}}
\newcommand{\tc}{\textnormal{TC}}
\newcommand{\fj}{\textnormal{FJ}}
\newcommand{\fc}{\textnormal{FC}}
\newcommand{\pcuts}{\textnormal{PC}}
\newcommand{\pjoins}{\textnormal{PJ}}
\newcommand{\rcuts}{\textnormal{RC}}
\newcommand{\rjoins}{\textnormal{RJ}}

\def\triangletr[#1](#2,#3){
	\draw[draw=black!30!white,#1] (#2+1,-#3) -- (#2+1,-#3-1) -- (#2,-#3-1) -- (#2+1,-#3);
}
\def\trianglebl[#1](#2,#3){
	\draw[draw=black!30!white,#1] (#2,-#3) -- (#2+1,-#3) -- (#2,-#3-1) -- (#2,-#3);
}
\def\cell[#1](#2,#3,#4,#5){
	\triangletr[fill=blue!#2!white](#4,#5);
	\trianglebl[fill=red!#3!white](#4,#5);
}

\begin{document}

\title{\bf Correlation Clustering of Bird Sounds}
\author{David Stein $\quad$ Bjoern Andres}
\date{TU Dresden}
\maketitle            

\begin{abstract}
Bird sound classification is the task of relating any sound recording to those species of bird that can be heard in the recording. Here, we study bird sound clustering, the task of deciding for any pair of sound recordings whether the same species of bird can be heard in both. We address this problem by first learning, from a training set, probabilities of pairs of recordings being related in this way, and then inferring a maximally probable partition of a test set by correlation clustering. We address the following questions: How accurate is this clustering, compared to a classification of the test set? How do the clusters thus inferred relate to the clusters obtained by classification? How accurate is this clustering when applied to recordings of bird species not heard during training? How effective is this clustering in separating, from bird sounds, environmental noise not heard during training?
\end{abstract}

\section{Introduction}

The abundance and variety of bird species are well-established markers of biodiversity and the overall health of ecosystems 
\cite{fitzpatrick2016handbook}. 
Traditional approaches to measuring these quantities rely on human experts counting bird species at select locations by sighting and hearing 
\cite{ralph1995monitoring}. 
This approach is labor-intensive, costly and biased by the experience of individual experts.
Recently, progress has been made toward replacing this approach by a combination of passive audio monitoring 
\cite{shonfield2017autonomous,Darras2018,MarkovaNenova2023}
and automated bird sound classification 
\cite{kahl2017large}.
The effectiveness of this automated approach can be seen, for instance, in \cite{Kahl2021,Wood2019}.
Bird sound classification is the task of relating any sound recording to those species of bird that can be heard in the recording
\cite{gupta-2021,kahl2017large}.
Models and algorithms for bird sound classification are a topic of the annual BirdCLEF Challenge
\cite{Goeau2018,kahl2020overview,kahl2021overview,kahl2022overview,kahl2019overview}.
Any model for bird sound classification is defined and learned for a fixed set of bird species.
At the time of writing, the most accurate models developed for this task all have the form of a neural network \cite{Goeau2018,gupta-2021,kahl2017acoustic,kahl2019overview,kahl2017large,Kahl2021,Salamon2017,sevilla2017}.

Here, we study bird sound clustering, the task of deciding for any pair of bird sound recordings whether the same species of bird can be heard in both.
We address this task in three steps.
Firstly, we define a probabilistic model of bird sound clusterings.
Secondly, we learn from a training set a probability mass function of the probability of pairs of sound recordings being related.
Thirdly, we infer a maximally probable partition of a test set by solving a correlation clustering problem locally.
Unlike models for bird sound classification, the model we define for bird sound clustering is agnostic to the notion of bird species.

In this article, we make four contributions:
Firstly, we quantify empirically how accurate bird sound correlation clustering is compared to bird sound classification.
To this end, we compare in terms of a metric known as the variation of information \cite{arabie-1973,meila-2007} partitions of a test set inferred using our model to partitions of the same test set induced by classifications of this set according to a fixed set of bird species.
Secondly, we measure empirically how the clusters of the test set inferred using our model relate to bird species.
To this end, we relate each cluster to an optimally matching bird species and count, for each bird species, the numbers of false positives and false negatives.
Thirdly, we quantify empirically how accurate correlation clustering is when applied to recordings of bird species not heard during training.
Fourthly, we quantify empirically the effectiveness of correlation clustering in separating from bird sounds environmental noise not heard during training.


\section{Related Work}
\label{section:related-work}

\emph{Metric-based clustering} of bird sounds with prior knowledge of the number of clusters is studied in \cite{Clementino2020UsingTL,McGinn2023,Seth2018}: $k$ means clustering in \cite{Seth2018}, $k$ nearest neighbor clustering in \cite{Clementino2020UsingTL}, and clustering with respect to the distance to all elements of three given clusters in \cite[Section 2.2]{McGinn2023}.
In contrast, we study \emph{correlation clustering} \cite{bansal-2004} of bird sounds without prior knowledge of the number of clusters.

In \cite{Clementino2020UsingTL}, the coefficients in the objective function of a clustering problem are defined by the output of a Siamese network.
Siamese networks, introduced in \cite{Bromley1993} and described in the recent survey \cite{Li2022}, are applied to the tasks of classifying and embedding bird sounds in \cite{Clementino2020UsingTL,Renter2021}.
We follow \cite{Clementino2020UsingTL} in that we also define the coefficients in the objective function of a clustering problem by the output of a Siamese network.
However, as we consider a correlation clustering problem, we learn the Siamese network by minimizing a loss function fundamentally different from that in \cite{Clementino2020UsingTL}.
Beyond bird sounds, correlation clustering with respect to costs defined by the output of a Siamese network is considered in \cite{ho2021learning,levinkov-2017-a,song2019end} for the task of clustering images, and in \cite{tang-2017-multiple} for the task of tracking humans in a video.
We are unaware of prior work on correlation clustering of bird sounds.

Probabilistic models of the partitions of a set, and, more generally, the decompositions of graphs, without priors or constraints on the number or size of clusters, are studied for various applications, including
image segmentation \cite{andres-2011,andres-2012-globally,kappes-2016,keuper-2015a},
motion trajectory segmentation \cite{keuper-2015b}
and multiple object tracking \cite{tang-2015,tang-2017-multiple}.
The Bayesian network we introduce here for bird sound clustering is analogous to the specialization to complete graphs of the model introduced in \cite{andres-2012-globally} for image segmentation.
Like in \cite{tang-2017-multiple} and unlike in \cite{andres-2012-globally}, the probability mass function we consider here for the probability of a pair of bird sounds being in the same cluster has the form of a Siamese network.
Like in \cite{andres-2012-globally} and unlike in \cite{tang-2017-multiple}, we cluster all elements, without the possibility of choosing a subset.

Complementary to prior work and ours on either classification or clustering of bird sounds are models for sound separation \cite{Wisdom2020} that can separate multiple bird species audible in the same sound recoding and have been shown to increase the accuracy of bird sound classification \cite{Denton2022}.

General theoretical connections between clustering and classification are established in \cite{bao2022,Zhang2007}.
\section{Model}
\label{section:model}

\begin{figure}[t]
\centering
\begin{tikzpicture}
\small
\node[draw=black, circle, label=right:$\mathcal{X}_{\{a,a'\}}$] at (1, 2) (x) {};
\node[draw=black, circle, label=right:$\mathcal{Y}_{\{a,a'\}}$] at (1, 1) (y) {};
\node[draw=black, circle, label=right:$\mathcal{Z}$] at (1, 0) (z) {};
\node[draw=black, circle, label=left:$\Theta_j$] at (0, 1) (theta) {};
\draw[-latex] (x) -- (y);
\draw[-latex] (theta) -- (y);
\draw[-latex] (y) -- (z);
\draw (-3.1, 0.5) rectangle (0.4, 1.5);
\draw (0.6, 0.5) rectangle (4.45, 2.5);
\node at (-2, 0.75) {$j \in \{1,\dots,n\}$};
\node at (3.5, 0.75) {$\{a,a'\} \in \tbinom{A}{2}$};
\end{tikzpicture}
\caption{Depicted above is a Bayesian network defining conditional independence assumptions of a probabilistic model for bird sound clustering we introduce in \Cref{section:bayesian-model}.}
\label{figure:bayesian-net}
\end{figure}
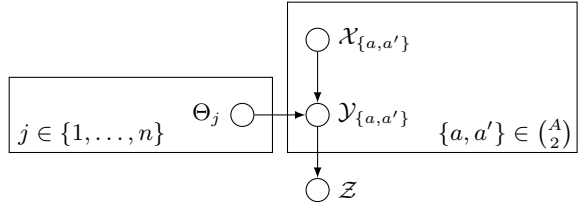

\subsection{Representation of clusterings}

We consider a finite, non-empty set $A$ of sound recordings that we seek to cluster.
The feasible solutions to this task are the partitions of the set $A$.
Recall that a partition $\Pi$ of $A$ is a collection $\Pi \subseteq 2^A$ of non-empty and pairwise disjoint subsets of $A$ whose union is $A$. Here, $2^A$ denotes the power set of $A$.
We will use the terms \emph{partition} and \emph{clustering} synonymously for the purpose of this article and refer to the elements of a partition as \emph{clusters}.

Below, we represent any partition $\Pi$ of the set $A$, by the function $y^\Pi: \tbinom{A}{2} \to \{0,1\}$ that maps any pair $\{a,a'\} \in \tbinom{A}{2}$ of distinct sound recordings $a,a' \in A$ to the number $y^\Pi_{\{a,a'\}} = 1$ if $a$ and $a'$ are in the same cluster, i.e.~if there exists a cluster $U \in \Pi$ such that $a \in U$ and $a' \in U$, and maps the pair to the number $y^\Pi_{\{a,a'\}} = 0$, otherwise.

Importantly, not every function $y: \tbinom{A}{2} \to \{0,1\}$ well-defines a partition of the set $A$. 
Instead, there can be three distinct elements $a,b,c$ such that $y_{\{a,b\}} = y_{\{b,c\}} = 1$ and $y_{\{a,c\}} = 0$.
However, it is impossible to put $a$ and $b$ in the same cluster, and put $b$ and $c$ in the same cluster, and not put $a$ and $c$ in the same cluster, as this violates transitivity.
The functions $y: \tbinom{A}{2} \to \{0,1\}$ that well-define a partition of the set $A$ are precisely those that hold the additional property
\begin{align}
\forall a \in A \;
\forall b \in A \setminus \{a\} \;
\forall c \in A \setminus \{a,b\} \colon \quad
	y_{\{a,b\}} + y_{\{b,c\}} - 1 \leq y_{\{a,c\}}
\enspace .
\label{eq:triangle-inequalities}
\end{align}
We let $Z_A$ denote the set of all such functions. 
That is:
\begin{align}
Z_A := \left\{
y^\Pi: \tbinom{A}{2} \to \{0,1\}
\ \middle|\ 
\eqref{eq:triangle-inequalities}
\right\}
\enspace .
\label{eq:feasible-set}
\end{align}

\subsection{Bayesian model}
\label{section:bayesian-model}

With the above representation of clusterings in mind, we define a probabilistic model with four classes of random variables. 
This model is depicted in \Cref{figure:bayesian-net}.

For every $\{a,a'\} \in \tbinom{A}{2}$, let $\mathcal{X}_{\{a,a'\}}$ be a random variable whose value is a vector $x_{\{a,a'\}} \in \mathbb{R}^{2m}$, with $m \in \mathbb{N}$.
We call the first $m$ coordinates a \emph{feature vector} of the sound recording $a$, and we call the last $m$ coordinates a feature vector of the sound recording $a'$.
These feature vectors are described in more detail in \Cref{section:experiments}.

For every $\{a,a'\} \in \tbinom{A}{2}$, let $\mathcal{Y}_{\{a,a'\}}$ be a random variable whose value is a binary number $y_{\{a,a'\}} \in \{0,1\}$, indicating whether the recordings $a$ and $a'$ are in the same cluster, $y_{\{a,a'\}} = 1$, or distinct clusters, $y_{\{a,a'\}} = 0$.

For a fixed number $n \in \mathbb{N}$ and every $j \in \{1, \dots, n\}$, let $\Theta_j$ be a random variable whose value is a real number $\theta_j \in \mathbb{R}$ that we call a \emph{model parameter}.

\vspace{-1.3ex} 
Finally, let $\mathcal{Z}$ be a random variable whose value is a set $Z \subseteq \{0,1\}^{\tbinom{A}{2}}$ of feasible maps from the set $\tbinom{A}{2}$ of pairs of distinct sound recordings to the binary numbers.
We will fix this random variable to the set $Z_A$ defined in \eqref{eq:feasible-set} of those functions that well-define a partition of the set $A$.

Among these random variables, we assume conditional independencies according to the Bayesian Net depicted in \Cref{figure:bayesian-net}.
This implies the factorization:
\begin{align}
P(\mathcal{X}, \mathcal{Y}, \mathcal{Z}, \Theta) 
= 
P(\mathcal{Z} \mid \mathcal{Y}) \,
\prod_{\mathclap{\{a,a'\} \in \tbinom{A}{2}}} P(\mathcal{Y}_{\{a,a'\}} \mid \mathcal{X}_{\{a,a'\}}, \Theta) \,
\prod_{\mathclap{\{a,a'\} \in \tbinom{A}{2}}} P(\mathcal{X}_{\{a,a\}})
\prod_{j=1}^{2m} P(\Theta_j)
\end{align}

\input{figure-siamese-network}

For the conditional probabilities on the right-hand side, we define probability measures:

First is a probability mass function that assigns a probability mass of zero to all $y \notin Z$ and assigns equal and positive probability mass to all $y \in Z$.
For any $Z \subseteq \{0,1\}^{\tbinom{A}{2}}$ and any $y \in \{0,1\}^{\tbinom{A}{2}}$:
\begin{align}
p_{\mathcal{Z} \mid \mathcal{Y}}(Z, y) \propto \begin{cases}
	1 & \text{if\ } y \in Z \\
	0 & \text{otherwise}
\end{cases}
\enspace .
\label{eq:subset-distribution}
\end{align}
Recall that we fix $Z = Z_A$, i.e.~we assign positive and equal probability mass to those binary labelings of pairs of audio recordings that well-define a clustering of the set $A$.

Second is a logistic distribution: For any $\forall \{a,a'\} \in \tbinom{A}{2}$, any $x_{\{a,a'\}} \in \mathbb{R}^{2m}$ and any $\theta \in \mathbb{R}^n$:
\begin{align}
	p_{\mathcal{Y}_{\{a,a'\}} \mid \mathcal{X}_{\{a,a'\}}, \Theta}(1, x_{\{a,a'\}}, \theta)
	=
	\frac{1}{1 + 2^{-f_\theta\left(x_{\{a,a'\}}\right)}}
\enspace .
\label{eq:logistic-distribution}
\end{align}
Here, the function $f_\theta \colon \mathbb{R}^{2m} \to \mathbb{R}$ has the form of the Siamese neural network depicted in \Cref{figure:siamese-network-sketch}.

Third is a uniform distribution on a finite interval. 
For a fixed $\tau \in \mathbb{R}^+$, any $j \in \{1,\dots,n\}$ and any $\theta_j \in \mathbb{R}$:
\begin{align}
p_{\Theta_j}(\theta_j) \propto
\begin{cases}
	1 & \text{if\ } \theta_j \in [-\tau, \tau] \\
	0 & \text{otherwise}
\end{cases}
\enspace .
\label{eq:prior-distribution}
\end{align}

\section{Learning}
\label{section:learning}


Training data consists of 
(i) a set $A$ of sound recordings, 
(ii) for each sound recording $a \in A$, a feature vector $x_a$, 
(iii) for each pair $\{a,a'\} \in \tbinom{A}{2}$ of distinct sound recordings, a binary number $y_{\{a,a'\}} \in \{0, 1\}$ that is $1$ if and only if a human annotator has labeled both $a$ and $a'$ with the same bird species.
This training data fixes the values of the random variables $X$ and $Y$ in the probabilistic model.
In addition, we fix $Z = Z_A$, as described above.

We learn model parameters by maximizing the conditional probability
\begin{align}
P(\Theta \mid \mathcal{X}, \mathcal{Y}, \mathcal{Z})
& \propto  \prod_{\{a,a'\} \in \tbinom{A}{2}} \hspace{-3ex} P(\mathcal{Y}_{\{a,a'\}} \mid \mathcal{X}_{\{a,a'\}}, \Theta) \prod_{j=1}^{2m} P(\Theta_j)
\enspace .
\end{align}

With the logistic distribution \eqref{eq:logistic-distribution} and the prior distribution \eqref{eq:prior-distribution}, and after elementary arithmetic transformations, this problem takes the form of the linearly constrained non-linear logistic regression problem
\begin{align}
\inf_{\theta \in \mathbb{R}^{2m}} \quad
	& \sum_{\{a,a'\} \in \tbinom{A}{2}} \left(
		- y_{\{a,a'\}} f_\theta(x_{\{a,a'\}}) + \log_2(1 + 2^{f_\theta(x_{\{a,a'\}})})
	\right)
\label{eq:learning-objective}
\\
\text{subject to} \quad
	& \forall j \in \{1,\dots,n\} \colon \ -\tau \leq \theta_j \leq \tau
\enspace .
\label{eq:learning-constraints}
\end{align}

In practice, we choose $\tau$ large enough for the constraints \eqref{eq:learning-constraints} to be inactive for the training data we consider, i.e.~we consider an uninformative prior over the model parameters.
We observe that the unconstrained problem \eqref{eq:learning-objective} is non-convex, due to the non-convexity of $f_\theta$.
In practice, we do not solve this problem, not even locally.
Instead, we compute a feasible solution $\hat\theta \in \mathbb{R}^n$ heuristically, by means of stochastic gradient descent with an adaptive learning rate.
More specifically, we employ the algorithm AdamW~\cite{loshchilov2018decoupled} with mini-batches $B_A \subseteq \tbinom{A}{2}$ and the loss 
\begin{equation}
	\frac{1}{\vert B_A\vert}\sum_{\{a, a'\}\in B_A}\left(-y_{\{a, a'\}}f_{\theta}(x_{\{a, a'\}}) + \log_2(1 + 2^{f_\theta(x_{\{a, a'\}})})\right)
\enspace.
\end{equation}

We set the initial learning rate to $10^{-4}$, the batch size to 64, and the number of iterations to 380,000. 
Moreover, we balance the batches in the sense that there are exactly $\vert B_A\vert / 2$ elements in $B_A$ with $y_{\{a, a'\}} = 1$ and exactly $\vert B_A\vert / 2$ elements in $B_A$ with $y_{\{a, a'\}} = 0$. 
All learning is carried out on a single NVIDIA A100 GPU with 16 AMD EPYC 7352 CPU cores, equipped with 32~GB of RAM.

\section{Inference}
\label{section:inference}


We assume to have learned and now fixed model parameters $\hat\theta$.
In addition, we are given a feature vector $x_a$ for every sound recording $a \in A$ of a test set $A$.
This fixes the values of the random variables $\Theta$ and $X$ in the probabilistic model.
In addition, we fix $Z = Z_A$, as described above, so as to concentrate the probability measure on those binary decisions for pairs of recordings that well-define a partition of the set $A$. 

We infer a clustering of the set $A$ by maximizing the conditional probability
\begin{align}
P(\mathcal{Y} \mid \mathcal{X}, \mathcal{Z}, \Theta) 
\ \propto \ 
P(\mathcal{Z} \mid \mathcal{Y}) \ 
\prod_{\mathclap{\{a,a'\} \in \tbinom{A}{2}}} P(\mathcal{Y}_{\{a,a'\}} \mid \mathcal{X}_{\{a,a'\}}, \Theta)
\end{align}

For the uniform distribution \eqref{eq:subset-distribution} on the subset $Z_A$, and for the logistic distribution \eqref{eq:logistic-distribution}, the maximizers of this probability mass can be found by solving the correlation clustering problem
\begin{align}
\max_{y \colon \tbinom{A}{2} \to \{0,1\}} \quad
& f_\theta(x_{\{a,a'\}}) \; y_{\{a,a'\}}
\\
\text{subject to} \quad
& \forall a \in A \;
\forall b \in A \setminus \{a\} \;
\forall c \in A \setminus \{a,b\} \colon \ 
	y_{\{a,b\}} + y_{\{b,c\}} - 1 \leq y_{\{a,c\}}
\end{align}

In practice, we compute a locally optimal feasible solution $\hat y \colon \tbinom{A}{2} \to \{0,1\}$ to this \textsc{np}-hard problem by means of the local search algorithm GAEC, until convergence, and then the local search algorithm KLj, both from 
\cite{keuper-2015a}.
The output $\hat y$ is guaranteed to well-define a clustering of the set $A$ such that any distinct sound recordings $a, a' \in A$ belong to the same cluster if and only if $\hat y_{\{a,a'\}} = 1$.

\section{Experiments}
\label{section:experiments}

\subsection{Dataset}
We start from those 17,313 audio recordings of a total of 316 bird species from the collection Xeno-Canto \cite{xeno-canto} of quality A or B that are recorded in Germany, contain bird songs and do not contain background species.
The files are re-sampled to 44,100~Hz and split into chunks of 2~seconds.
For each chunk, we compute the mel spectrogram with a frame width of 1024 samples, an overlap of 768 samples and 128 mel bins and re-scale it to $128\times384$ entries.
Finally, to distinguish salient from non-salient chunks, we apply the signal detector proposed in \cite{Kahl2021}. 
Bird species with less than 100 salient audio chunks are excludeed.
This defines a first dataset of 68 bird species with at least 10 minutes of audio recordings in total. 
We split this set according to the proportions 8/1/1 into disjoint subsets Train-68, Val-68 and Test-68.
In addition, we consider a set Test-0,87 of 87 bird species with less than 10 minutes but more than one minute of audio data.
We call the union of both test sets Test-68,87.
In addition, we define a set Test-N containing 39 classes of environmental noise not used for augmentation from the collection ESC-50 \cite{piczak2015dataset}.
We refer to the union of Test-68 and Test-N as Test-68,N.
During learning, we employ augmentation techniques, specifically: horizontal and vertical roll, time shift, SpecAugment \cite{Park19}, as well as the addition of white noise, pink noise and some environmental noise from ESC-50.

	\subsection{Metrics}

In order to measure the distance between a predicted partition $\hat\Pi$ of a finite set $A$, on the one hand, and a true partition $\Pi$ of the same set $A$, on the other hand, we evaluate a metric known as the variation of information \cite{arabie-1973,meila-2007}:
\begin{equation}
	\vi(\Pi, \hat\Pi) = H(\Pi \mid \hat\Pi) + H(\hat\Pi \mid \Pi)
\end{equation}
Here, the conditional entropy $H(\Pi \mid \hat\Pi)$ is indicative of false joins, whereas the conditional entropy $H(\hat\Pi \mid \Pi)$ is indicative of false cuts.

In order to measure the accuracy of decisions $\hat{y}\colon \tbinom{A}{2}\to \{0, 1\}$ for all pairs $\{a,a'\} \in \tbinom{A}{2}$ of sound recordings also for decisions that do not well-define a clustering of $A$, we calculate the numbers of true joins (TJ), true cuts (TC), false cuts (FC) and false joins (FJ) of these pairs according to \Cref{eq:tjtc,eq:fcfj} below.
From these, we calculate in the usual way the precision and recall of cuts, the precision and recall of joins, and Rand's index \cite{rand-1971}.
\begin{align}
	\tj(y^\Pi, \hat{y}) = \sum_{ij\in \tbinom A2} y^\Pi_{ij} \hat{y}_{ij} \enspace, \quad &
	\tc(y^\Pi, \hat{y}) = \sum_{ij\in \tbinom A2} (1-y^\Pi_{ij})(1 - \hat{y}_{ij}) 
	\label{eq:tjtc}
	\\
	\fc(y^\Pi, \hat{y}) = \sum_{ij\in \tbinom A2} (1-\hat{y}_{ij})y^\Pi_{ij}\enspace, \quad &
	\fj(y^\Pi, \hat{y}) = \sum_{ij\in \tbinom A2} \hat{y}_{ij} (1 - y^\Pi_{ij}) \enspace.
	\label{eq:fcfj}
\end{align}

	\subsection{Clustering vs Classification}
\label{section:clustering-vs-classification}

\begin{table}[!b]
	\centering
	\scriptsize
	\begin{tabular}{rlrrrrrrrrrr}
		\toprule 
		& Model  & $\Pi$ & RI & $\vi$ & $\vi_{\textnormal {FC}}$ & $\vi_{\textnormal{FJ}}$ & PC & RC & PJ & RJ & CA \\
		\midrule
		1. & $f_\theta$ & no & 0.89 & - & - & - & 97.9\% & 89.9\% & 42.6\% & 79.5\% & - 
		\\
		2. &$f_\theta$ + Aug & no & 0.87 & - & - & - & \textbf{98.7\%} & 86.9\% & 38.7\% & \textbf{87.6\%} & - 
		\\
		\midrule
		3. &$f_\theta$ + CC & yes & 0.93 & 4.21 & 1.99 & 2.22 & 97.3\% & 95.0\% & 57.6\% & 72.1\% & - 
		\\
		4. &$f_\theta$ + Aug + CC & yes & 0.91 & 3.28 & 1.34 & 1.95 & 98.1\% & 92.0\% & 48.9\% & 81.3\% & - 
		\\
		5. &$f_\theta$ + CC + T & yes & 0.93 & 4.21 & 2.02 & 2.19 & 97.3\% & 95.2\% & 58.5\% & 71.7\% & - 
		\\
		6. &$f_\theta$ + Aug + CC + T & yes & 0.91 & 3.27 & 1.35 & 1.91 & 98.1\% & 92.2\% & 49.4\% & 80.8\% & - 
		\\
		\midrule
		7. & ResNet18 & yes & 0.94 & 4.67 & 2.33 & 2.34 & 96.7\% & 96.9\% & 66.2\% & 64.8\% & 59.6\% 
		\\
		8. &ResNet18 + Aug & yes & \textbf{0.96} & \textbf{3.20} & 1.68 & \textbf{1.72} & 97.3\% & \textbf{97.8\%} & \textbf{75.3\%} & 71.8\% & 72.7\% 
		\\
		9. &BirdNET Analyzer & yes & 0.77 & 3.50 & \textbf{1.22} & 2.28 & 94.3\% & 79.4\% & 18.3\%& 48.9\% & 49.7\% 
		\\
		10. &$f_\theta$ + T & yes & 0.93 & 4.26 & 1.97 & 2.29 & 97.3\% & 95.0\% & 57.8\% & 72.2\% & 64.1\% 
		\\
		11. &$f_\theta$ + Aug + T & yes & 0.94& 3.31 & 1.48 & 1.83 & 97.8\% & 95.6\% & 62.6\% & 77.3\% &  \textbf{73.1\%} 
		\\
		\bottomrule
	\end{tabular}
	\\[1ex]
	\caption{Above, we report, for models trained on Train-68 and evaluated on Test-68, whether the inferred solution well-defines a partition of Test-68 ($\Pi$) and how this solution compares to the truth in terms of Rand's index (\ri), the variation of information (\vi), conditional entropies due to false cuts ($\vi_{\textnormal{FC}}$) and false joins ($\vi_{\textnormal{FJ}}$), the precision (P) and recall (R) of cuts (C) and joins (J), and the classification accuracy (CA).}
	\label{table:birds-69-test-dataset-with-classification-metrics}
\end{table}

Here, we describe the experiments we conduct in order to compare the accuracy of a clustering of bird sounds with the accuracy of a classification of bird sounds.
The results are shown in \Cref{table:birds-69-test-dataset-with-classification-metrics} and \Cref{figure:truth-prediction-cluster-matrix-sorted}.

\textbf{Procedure and results.}
Toward clustering, we learn the model $f_\theta$ defined in \Cref{section:bayesian-model}, as described in \Cref{section:learning}, from the data set Train-68, with and without data augmentation, and apply it to the independent data set Test-68 in two different ways:
Firstly, we infer an independent decision $y_{\{a,a'\}} \in \{0,1\}$ for every pair of distinct sound recordings $a, a'$, by asking whether $f_\theta(x_{\{a,a'\}}) \geq 0$ ($y_{\{a, a'\}} = 1$) or $f_{\theta}(x_{\{a, a'\}}) < 0$ ($y_{\{a, a'\}} = 0$). 
These decisions together do not necessarily well-define a clustering of Test-68.
Yet, we compare these decisions independently to the truth, in Rows 1-2 of \Cref{table:birds-69-test-dataset-with-classification-metrics}.
Secondly, we infer a partition of Test-68 by correlation clustering, as described in \Cref{section:inference} (Rows 3-4 of \Cref{table:birds-69-test-dataset-with-classification-metrics}).
Thirdly, we infer a partition of Test-68 and a subsample of Train-68, which contains 128 randomly chosen recordings per species, jointly by locally solving the correlation clustering problem for the union of these data sets, also as described in \Cref{section:inference}; (Rows 5-6 of \Cref{table:birds-69-test-dataset-with-classification-metrics}).

Toward classification, we learn a ResNet-18 on Train-68, with and without data augmentation.
Using this model, we infer a classification of Test-68 (Rows 7-8 of \Cref{table:birds-69-test-dataset-with-classification-metrics}).
In addition, we classify Test-68 by means of BirdNET analyzer~\cite{Kahl2021} (Row 9 of \Cref{table:birds-69-test-dataset-with-classification-metrics}).
We remark that BirdNET is defined for 3-second sound recordings while we work with 2-second sound recordings. 
When applying BirdNET to these 2-second recordings, they are padded with random noise as described in \cite{birdnet-analyzer}.
Finally, we infer a classification of Test-68 by assigning each sound recording to one of the true clusters of Train-68 for which this assignment is maximally probable according to the model $f_\theta$ learned on Train-68.
We report the accuracy of this classification with respect to $f_\theta$ in Rows 10-11 of \Cref{table:birds-69-test-dataset-with-classification-metrics}.
For each classification of Test-68, we report the distance from the truth of the \emph{clustering} of Test-68 induced by the classification. 
This allows for a direct comparison of classification with clustering.

\textbf{Discussion.}
Closest to the truth by a variation of information of $3.20$ is the clustering of Test-68 induced by the classification of Test-68 by means of the ResNet-18 learned from Train-68, with data augmentation (Row 8 in \Cref{table:birds-69-test-dataset-with-classification-metrics}).
This result is expected, as classification is clustering with a constrained set of clusters, and this constraint constitutes additional prior knowledge.
Dropping this information during learning but not during inference (Row 6 in \Cref{table:birds-69-test-dataset-with-classification-metrics}) leads to the second best clustering that differs from the true clustering of Test-68 by a variation of information of $3.27$.
Dropping this knowledge during learning and inference (Row 4 in \Cref{table:birds-69-test-dataset-with-classification-metrics}) leads to a variation of information $3.28$.
It can be seen from these results that a clustering of this bird sound data set is less accurate than a classification, but still informative.
From a comparison of Rows 2 and 4 of \Cref{table:birds-69-test-dataset-with-classification-metrics}, it can bee seen that the local solution of the correlation clustering problem not only leads to decisions for pairs of sound recordings that well-define a clustering of Test-68 but also increases the accuracy of these decisions in terms of Rand's index, from 0.87 to 0.91.
Looking at these two experiments in more detail, we observe an increase in the recall of cuts and precision of joins due to correlation clustering, while the precision of cuts decreases slightly and the recall of joins decreases strongly.
Indeed, we observe more clusters than bird species (see \Cref{figure:truth-prediction-cluster-matrix-sorted}).
There are two possible explanations for this effect.
Firstly, the local search algorithm we apply starts from the finest possible clustering into singleton sets and is therefore biased toward excessive cuts (more clusters).
Secondly, there might be different types of sounds associated with the same bird species.
We have not been able to confirm or refute this hypothesis and are encouraged to collaborate with ornithologists to gain additional insight.

	\subsection{Clustering Unseen Data}
\label{section:experiment-unseen}
\begin{table}[!b]
	\centering
	\scriptsize
	\begin{tabular}{rlrrrrrrrrrr}
		\toprule
		& Model &  $\Pi$ & \ri & $\vi$ & $\vi_{\textnormal {FC}}$ & $\vi_{\textnormal{FJ}}$ & \pcuts & \rcuts & \pjoins & \rjoins & CA \\
		\midrule
		1. & $f_\theta$ & no & 0.82 & - & - & -  & 97.5\% & 83.5\% & 14.6\% & 57.1\% & -
		\\
		2. & $f_\theta$ + Aug & no & 0.78 & - & - & - & \textbf{97.8\%} & 79.2\% & 13.1\% & \textbf{64.0\%} & - 
		\\
		\midrule
		3. & $f_\theta$ + CC & yes & \textbf{0.90} & 5.42 & 2.30 & \textbf{3.12} & 96.9\% & \textbf{92.4\%} & \textbf{20.8\%} & 40.9\% & 37.7\% 
		\\
		4. & $f_\theta$ + Aug + CC & yes & 0.86 & \textbf{5.06} & \textbf{1.83} & 3.23 & 97.2\% & 88.4\% & 16.7\% & 47.5\% & 39.4\% 
		\\
		\bottomrule
	\end{tabular}
	\caption{Above, we report the accuracy of the learned model $f_\theta$ when applied to the task of clustering the data set Test-0,87 of bird sounds of 87 bird species not heard during training.}
	\label{table:birds-68-unseen-birds-only}
\end{table}
\begin{table}[!b]
	\centering
	\scriptsize
	\begin{tabular}{rlrrrrrrrr}
		\toprule 
		& Model & $\Pi$ &  & 
		$J_{\textnormal{UU}}$ & $C_{\textnormal{UU}}$ & $J_{\textnormal{UB}}$ & $C_{\textnormal{UB}}$ & $J_{\textnormal{BB}}$ & $C_{\textnormal{BB}}$ \\
		\midrule
		1. & $f_\theta$ & no & P: & 14.6\% & 97.5\% & 0\% & 100\% & 42.6\% & 97.9\%
		\\
		& & & R: & 57.1\% & 83.5\% & 100\% & 84.6\% & 79.5\% & 89.9\%
		\\
		2. & $f_\theta$ + Aug & no & P: & 13.1\% & \textbf{97.8\%} & 0\% & 100\% & 38.7\% & \textbf{98.7\%}
		\\
		& & & R: & \textbf{64.0\%} & 79.2\% & 100\% & 81.1\% & \textbf{87.6\%} & 86.9\%
		\\
		\midrule
		3. & $f_\theta$ + CC & yes & P: & 14.3\% & 96.1\% & 0\% & 100\% & \textbf{59.7\%} & 97.1\%
		\\
		& & & R: & 23.2\% & \textbf{93.2\%} & 100\% & \textbf{91.7\%} & 70.1\% & \textbf{95.5\%}
		\\
		4. & $f_\theta$ + CC + Aug & yes & P: & \textbf{17.7\%} & 96.8\% & 0\% & 100\% & 47.7\% & 98.1\%
		\\
		& & & R: & 39.0\% & 91.1\% & 100\% & 89.0\% & 81.3\% & 91.6\%
		\\
		\bottomrule
	\end{tabular}
	\caption{Above, we report the accuracy of the learned model $f_\theta$ when applied to the task of clustering the data set Test-68,87 of bird sounds of 68 bird species heard during training and 87 bird species not heard during training.
	More specifically, we report precision and recall of cuts and joins, separately for pairs of sound recordings both belonging to Test-0,87 (UU), both belonging to Test-68 (BB) or containing one from the set Test-0,87 and one from the set Test-68 (UB).}
	\label{table:birds-68-unseen-birds-and-test}
\end{table}
Next, we describe the experiments we conduct in order to quantify the accuracy of the learned model for bird sound clustering when applied to sounds of bird species not heard during training.
The results are shown in
\Cref{table:birds-68-unseen-birds-only}.
Additional results for a combination of bird species heard and not heard during training are shown in \Cref{table:birds-68-unseen-birds-and-test}.

\textbf{Procedure and results.}
To begin with, we learn $f_\theta$ on Train-68 as described in \Cref{section:learning}.
Then, analogously to \Cref{section:clustering-vs-classification}, we infer an independent decision $y_{\{a,a'\}} \in \{0,1\}$ for every pair of distinct sound recordings $a, a'$ from the data set Test-0,87, by asking whether $f_\theta(x_{\{a,a'\}}) \geq 0$.
We compare these independent decisions to the truth, in Rows 1-2 of \Cref{table:birds-68-unseen-birds-only}.
Next, we infer a partition of Test-0,87 by correlation clustering, as described in \Cref{section:inference}; see Rows 3-4 of \Cref{table:birds-68-unseen-birds-only}.
Analogously to these two experiments, we infer decisions and a partition of the joint test set Test-68,87; see \Cref{table:birds-68-unseen-birds-and-test}.

\textbf{Discussion.}
It can be seen from Rows 3 and 4 of \Cref{table:birds-68-unseen-birds-only} that a clustering inferred using the model $f_\theta$ of the bird sounds of the data set Test-0,87 of 87 bird species not contained in the training data Train-68 is informative, i.e.~better than random guessing.
Furthermore, it can be seen from a comparison of Rows 1 and 3 as well as from a comparison of Rows 2 and 4 of \Cref{table:birds-68-unseen-birds-only} that correlation clustering increases the recall of cuts and the precision of joins, but decreases the precision of cuts and the recall of joins. 
Precision and recall of cuts are consistently higher than precision and recall of joins.
This observation is consistent with the excessive cuts we have observed also for bird species seen during training, cf.~Section~\ref{section:clustering-vs-classification}.
Possible explanations are, firstly, the bias toward excessive cuts in clusterings output by the local search algorithm we use for the correlation clustering problem and, secondly, the presence of different types of sounds for the same bird species in the data set Test-0,87.
From \Cref{table:birds-68-unseen-birds-and-test}, it can be seen that the clustering inferred using $f_\theta$ separates heard from unheard bird species accurately. 
From a comparison of \Cref{table:birds-69-test-dataset-with-classification-metrics,table:birds-68-unseen-birds-and-test,table:birds-68-unseen-birds-only}, it can be seen for pairs of bird sounds both from species heard during training (BB) or both from species not heard during training (UU), that the accuracy degrades little in a clustering of the joint set Test-68,87, compared to clusterings of the separate sets Test-68 and Test-0,87.
 
	\begin{table}[!b]
	\centering
	\scriptsize
	\begin{tabular}{lrrrrrrrr}
		\toprule 
		Model & $\Pi$ &  & 
		$J_{\textnormal{NN}}$ & $C_{\textnormal{NN}}$ & $J_{\textnormal{NB}}$ & $C_{\textnormal{NB}}$ & $J_{\textnormal{BB}}$ & $C_{\textnormal{BB}}$ \\
		\midrule
		$f_\theta$ & no & P: & \textbf{3.9\%} & 98.5\% & 0\% & \textbf{100\%} & 42.6\% & 97.9\%
		\\
		& & R: & 64.1\% & 59.2\% & 100\% & 78.4\% & 79.5\% & 89.9\%
		\\
		$f_\theta$ + Aug & no & P: & 3.2\% & \textbf{98.9\%} & 0\% & \textbf{100\%} & 38.7\% & \textbf{98.7\%}
		\\
		& & R: & \textbf{84.9\%} & 34.2\% & 100\% & 77.9\% & \textbf{87.6\%} & 86.9\%
		\\
		\midrule
		$f_\theta$ + CC & yes & P: & 3.3\% & 97.7\% & 0\% & \textbf{100\%} & \textbf{57.5\%} & 97.3\%
		\\
		& & R: & 25.0\% & \textbf{81.4\%} & 100\% & 87.0\% & 72.0\% & \textbf{95.0\%}
		\\
		$f_\theta$ + CC + Aug & yes & P: & 3.5\% & 98.0\% & 0\% & \textbf{100\%} & 47.7\% & 98.2\%
		\\
		& & R: & 47.5\% & 66.8\% & 100\% & \textbf{88.2\%} & 82.3\% & 91.5\%
		\\
		\bottomrule
	\end{tabular}
	\caption{Above, we report the accuracy of the learned model $f_\theta$ on Test-68,N. 
		This includes precision and recall of cuts and joins for pairs of recordings both the from Test-N (NN), both from Test-68 (BB) or one from Test-N and one from Test-68 (NB).}
	\label{table:birds-69-noise-and-test}
\end{table}
\subsection{Clustering Noise}
Next, we describe the experiments we conduct in order to quantify the accuracy of clusterings, inferred using the learned model, of bird sounds and environmental noise not heard during training.
The results are shown in
\Cref{table:birds-69-noise-and-test}.

\textbf{Procedure and results.}
To begin with, we learn $f_\theta$ on the data set Train-68 as described in \Cref{section:learning}.
Then, analogously to \Cref{section:experiment-unseen}, we infer an independent decision $y_{\{a,a'\}} \in \{0,1\}$ for every pair of distinct sound recordings $a, a'$ from the data set Test-68,N, by asking whether $f_\theta(x_{\{a,a'\}}) \geq 0$. 
We compare these independent decisions to the truth, in Rows 1-2 of \Cref{table:birds-69-noise-and-test}.
Next, we infer a partition of Test-68,N by correlation clustering, as described in \Cref{section:inference}; see Rows 3-4 of \Cref{table:birds-69-noise-and-test}.

\textbf{Discussion.}
From \Cref{table:birds-69-noise-and-test}, it can be seen that $f_\theta$ separates environmental noise form the set Test-N accurately from bird sounds from the set Test-68, with or without correlation clustering, and despite the fact that the noise has not been heard during training on Train-68.
From a comparison of \Cref{table:birds-69-noise-and-test,table:birds-69-test-dataset-with-classification-metrics}, it can be seen that the clustering of those sound recordings that both belong to Test-68 (BB) degrades only slightly when adding the environmental noise from the set Test-N to the problem.
From the column $J_{NB}$ and $C_{NB}$ of \Cref{table:birds-69-noise-and-test}, it can be seen that clustering the 39 types of noise is more challenging.
This is expected, as environmental noise is different from bird sounds and has not been heard during training.

\section{Conclusion}
\label{section:conclusion}

We have defined a probabilistic model, along with heuristics for learning and inference, for clustering sound recordings of birds by estimating for pairs of recordings whether the same species of bird can be heard in both.
For a public collection of bird sounds, we have shown empirically that partitions inferred by our model are less accurate than classifications with a known and fixed set of bird species, but are still informative.
Specifically, we have observed more clusters than bird species.
This observation encourages future work toward solving the instances of the inference problem exactly, with the goal of eliminating a bias toward additional clusters introduced by the inexact local search algorithm we employ here.
This observation also encourages future collaboration with ornithologists toward an analysis of the additional clusters.
Finally, our model has proven informative when applied to sound recordings of 87 bird species not heard during training, and in separating from bird sounds 39 types of environmental noise not used for training.
Further work is required to decide if this can be exploited in practice, e.g.~for rare species with little training data.

\subsection*{Acknowledgment}

The authors acknowledge funding by the Federal Ministry of Education and Research of Germany, from grant 01LC2006A.

\input{figure-heatmap-truth-predicted-sorted}

\bibliographystyle{plainurl}
\bibliography{manuscript}

\end{document}